\documentstyle[preprint,aps]{revtex}
\begin{document}
\draft
\title{
Level spacing distribution \\ 
of pseudointegrable billiard
}
\author{
T.Shigehara
}
\address{
Computer Centre, University of Tokyo, 
Bunkyo-ku Yayoi, Tokyo 113, Japan
}
\author{
N.Yoshinaga 
}
\address{
Department of Physics, College of Liberal Arts, 
Saitama University, 
Saitama 338, Japan
}
\author{
Taksu Cheon
}
\address{
Department of Physics, Hosei University, Fujimi, 
Chiyoda-ku, Tokyo 102, Japan
}
\author{
T. Mizusaki 
}
\address{
Department of Physics, University of Tokyo, 
Bunkyo-ku Hongo, Tokyo 113, Japan
}
\date{September 14, 1992}
\maketitle
\begin{abstract}
In this paper, we examine 
the level spacing distribution $P(S)$ 
of the rectangular billiard with a single point-like scatterer, 
which is known as pseudointegrable.
It is shown that 
the observed $P(S)$ is a new type, which is 
quite different from the previous conclusion.
Even in the strong coupling limit, 
the Poisson-like behavior rather than Wigner-like is seen 
for $S>1$, although the level repulsion still remains 
in the small $S$ region.
The difference from the previous works 
is analyzed in detail.
\end{abstract}
\pacs{05.45.+b}
%
%
%
%
\narrowtext
   The {\em wave chaos} has recently attracted much attention 
as one of typical manifestations of quantum chaos. 
The wave chaos is a new class of 
chaos which is generated 
by purely quantal effects in the systems, 
the classical counterpart of which is non-chaotic. 
Many numerical and theoretical works have already been done on the 
pseudointegrable dynamical systems with a few degrees of freedom 
[1-17]. 
One of such systems is the two-dimensional rectangular billiard with 
a single point-like scatterer, for which the statistical properties 
of the eigenvalues and wave functions have been examined 
from various points of view [12-14]. 
In the recent paper \cite{sz91}, 
\u{S}eba and \.{Z}yczkowski examined this system 
throughout a wide range of energy excitation 
using the Green's function method and revealed some new aspects. 
One of their conclusions is that the level spacing distribution $P(S)$ 
becomes closer to the Wigner distribution 
in the strong coupling limit, 
although the fine structure does not conform to the prediction of GOE 
(Gaussian Orthogonal Ensemble). 
The appearance of Wigner-like distribution in $P(S)$ might be considered
as further evidence 
which confirms the existence of the wave chaos.

There are now two approaches to the investigation of this 
pseudointegrable billiard. One of them is, as mentioned above, 
to analyze the Green's function of this system, and the other is 
to diagonalize the Hamiltonian by using the Fourier basis. 
While both approaches do not include any approximation formally, 
the truncation of the basis is inevitable 
in the actual numerical calculation. 
One should be most careful to decide the range of the basis 
because of the singularity of the interaction under consideration. 
In this brief paper, we will examine the effect of 
restriction of the basis on $P(S)$. 
This might sound strange because 
it might appear to be a mere technical problem 
in the numerical treatment lacking any physical interests. 
As we will see later, 
however, this is not the case and the closest care to the basis 
is essential for this system. 
In fact, the main conclusion is 
that the level spacing distribution $P(S)$ in the strong coupling limit 
never becomes Wigner-like but belongs to a quite new class.

More systematic analysis including the investigation of 
the other statistical properties 
than $P(S)$ concerning the eigenvalues and wave functions 
will be performed in the forthcoming paper \cite{sycm92}. 
Here, we will restrict ourselves to revealing the influence 
of the limitation of the phase space on $P(S)$.

We follow the Green's function method. As the mathematical formalism 
is explained in full in \cite{sz91}, we only 
summarize those points that are necessary in the following discussion, 
stressing the physical aspects. 
The Hamiltonian of the two-dimensional rectangular billiard 
with a point-like scatterer
is formally given by
\begin{equation}
\label{eq:e1}
  H = -\frac{\Delta}{2M}+v_{0}\delta(x-x_{0})\delta(y-y_{0}),
\end{equation}
where $M$ is the mass of a particle, $v_{0}$ and 
$(x_{0},y_{0})$ are strength and 
position of the scatterer respectively. 
The Green's function of this system is given by

\vspace{5mm}
  \hspace{1cm} $G(x,y;x',y';z)= 
  G^{(0)}_{F}(x,y;x',y';z)$

\vspace{5mm}
\begin{equation}
\label{eq:e1a}
  +
  G^{(0)}_{F}(x,y;x_{0},y_{0};z)
  \frac{v_{0}}{1-G^{(0)}(x_{0},y_{0};x_{0},y_{0};z)v_{0}}
  G^{(0)}_{F}(x_{0},y_{0};x',y';z).
\end{equation}
Here, $z$ is the energy variable,  
$G^{(0)}_{F}$ is the Green's function of the billiard 
without any scatterer and $G^{(0)}(x_{0},y_{0};x_{0},y_{0};z)$ 
describes 
the propagation of the particle which begins to propagate at the 
point-like scatterer and ends there. Clearly, the second 
term on the r.h.s. in Eq.(\ref{eq:e1a}) means the multiple 
scattering caused by the point-like scatterer.
(Although the authors of \cite{sz91} call 
$G^{(0)}(x_{0},y_{0};x_{0},y_{0};z)$ 
with opposite sign 
in Eq.(\ref{eq:e1a})
 the meromorphic function $\xi(z)$, we call it just 
the Green's function in the following discussion as long as causing 
no confusion.)
From Eq.(\ref{eq:e1a}), we see that, in the Green's function 
method, the eigenvalue problem is equivalent to 
solving the following equation, 
\begin{equation}
\label{eq:e2}
  G^{(0)}(x_{0},y_{0};x_{0},y_{0};z)=\frac{1}{v_{0}}.
\end{equation}
Owing to the fact that the obstacle is point-like, the whole problem 
can be reduced 
to a transcendental equation instead of an integral equation. 
If the scatterer is located at the center of the rectangle, which is 
the case we will examine in this paper, the Green's function 
(meromorphic function) with the Dirichlet condition 
on the border of the billiard 
is given by
\begin{equation}
\label{eq:e3}
  G^{(0)}(z)\equiv \\
  G^{(0)}(x_{0},y_{0};x_{0},y_{0};z)=
  \frac{4}{l_{x}l_{y}}
  \sum_{n_{x},n_{y}=1}^{\infty}\{
  \frac{1}{z-E_{2n_{x}-1,2n_{y}-1}^{(0)}}+
  \frac{E_{2n_{x}-1,2n_{y}-1}^{(0)}}
       {(E_{2n_{x}-1,2n_{y}-1}^{(0)})^{2}+1}\},
\end{equation}
\begin{equation}
\label{eq:e4}
  E_{n_{x},n_{y}}^{(0)}=\frac{\pi^{2}}{2M}\{(\frac{n_{x}}{l_{x}})^{2}+
  (\frac{n_{y}}{l_{y}})^{2}\}.
\end{equation}
Here, $E_{n_{x},n_{y}}^{(0)}$ is the eigenvalue of the billiard 
without any obstacle 
and $l_{x}$ and $l_{y}$ are the side-lengths  of the rectangle
 ($x_{0}=l_{x}/2$ and $y_{0}=l_{y}/2$). 
One should notice that, when the scatterer is placed at the 
center of the billiard, the scatterer affects only even-even parity 
states.
The special feature in case of the singular interaction can be seen 
in the second term in the Green's function of Eq.(\ref{eq:e3}), 
while the Green's function $G^{(0)}_{F}$ in the case 
without any obstacle 
in the billiard does not have the corresponding term. 
(The appearance of this term is closely related to 
the boundary condition around the scatterer. 
In order to determine its exact form, 
one needs the help of some theorems in the functional analysis. 
For details, see Ref.\cite{sz91} and references therein.) 
One realizes that each of two terms in the Green's function 
$G^{(0)}(z)$ has 
logarithmic divergence when summed separately, 
although the sum of them leads to a finite value.

We examine the case that $M=8\pi$, 
$l_{x}=\pi/3$ and $l_{y}=3/\pi$. 
In this particular parameterization, the average density of even-even 
parity states is equal to one according to the Weyl's formula. 

To see a general feature of the Green's function $G^{(0)}(z)$, 
the schematic graph 
of the Green's function is shown in Fig.1. 
Here, the eigenvalues $\{E_{2n_{x}-1,2n_{y}-1}^{(0)}\}$ 
of even-even parity states in the unperturbed 
system are renamed in ascending order as $\{E_{n}^{(0)}\}$. 
One can easily see that each eigenvalue $E_{n}$ 
of the perturbed system is isolated between two 
unperturbed energies, $E_{n}^{(0)}$ and $E_{n+1}^{(0)}$, namely
\[ E_{1}^{(0)}<E_{1}<E_{2}^{(0)}<E_{2}<E_{3}^{(0)}<E_{3}<\cdots,   \]
and becomes larger as one increases the strength of 
the coupling. 
In the strong coupling limit ($v_{0}=\infty$), 
the set of eigenvalues is just that of 
zeros of the Green's function. 

In order to get the solutions of Eq.(\ref{eq:e2}) numerically, 
one must limit the range of the summation with minimum and maximum 
values of $n$, $n_{min}$ and $n_{max}$,
\begin{equation}
\label{eq:e5}
  G^{(0)}_{appro}(z)=
  4\sum_{n=n_{min}}^{n_{max}}\{\frac{1}{z-E_{n}^{(0)}}+
  \frac{E_{n}^{(0)}}{(E_{n}^{(0)})^{2}+1}\}.
\end{equation}
The prescription of the limitation in \cite{sz91} 
is to take $n_{min}=l-500$ and 
$n_{max}=l+500$ for looking for a root $E_{l}$ of the Eq.(\ref{eq:e2}) 
localized between 
$E_{l}^{(0)}$ and $E_{l+1}^{(0)}$. 
Hereafter, we quote this prescription as the {\em truncation (I)}. 
At first sight, this seems to be quite reasonable because the main 
contribution on the Green's function around the energy $E_{l}$ comes 
from the terms that have $n$ around $l$ 
and because the contribution on the 
Green's function from $n<n_{min}$ tends to cancel that from 
$n>n_{max}$. 
According to the truncation (I), one gets the level spacing 
distribution $P(S)$ without much numerical labor. 
As a typical example, we show the case of the strong coupling limit 
in Fig.2. 
This corresponds to the Fig.2c in \cite{sz91} and of course shows quite 
similar structure, although the parameters for the system are slightly 
different in both calculations. 
One might conclude from Fig.2 that the level spacing distribution 
of the rectangular billiard with a point-like scatterer is almost 
Wigner-like in the strong coupling limit.

We now examine the accuracy of the truncation (I). 
Fig.3 shows the same calculation as above except that $n_{min}=1$ and 
$n_{max}=100000$. 
Hereafter, We quote this case as the {\em truncation (II)}. 
We have numerically checked sufficient convergence of the eigenvalues 
in the energy region under consideration. Also, we will later justify 
this truncation of the basis in an analytic manner in this paper. 
One easily sees the drastic changes even in a qualitative level. 
For $S>1$, 
$P(S)$ is rather Poisson-like than Wigner-like, 
although the level repulsion still remains in the small $S$ region. 
The level repulsion is regarded as a common feature 
among the various pseudointegrable systems \cite{l90,ss91,zs92}.
Roughly speaking, one might say that $P(S)$ in Fig.3 
shows an {\em intermediate} 
feature between the regularity and the chaos.

In order to clarify the reason for the disagreement between 
the level spacing distributions in
Fig.2 and Fig.3, 
we estimate the numerical error in the Green's 
function related to the truncation of the basis. 
The numerical error comes from the terms which are neglected by 
the limitation of summation in Eq.(\ref{eq:e5}),
\begin{equation}
\label{eq:e6}
  \delta G^{(0)}(z) \equiv G^{(0)}(z) -G^{(0)}_{appro}(z) \\
  = 4(\sum_{n=1}^{n_{min}-1}+\sum_{n=n_{max}+1}^{\infty})
  \{\frac{1}{z-E_{n}^{(0)}}+
  \frac{E_{n}^{(0)}}{(E_{n}^{(0)})^{2}+1}\}.
\end{equation}
To estimate the order of magnitude of error, 
we consider the unperturbed 
energy $E_{n}^{(0)}$ as the continuous variable and replace 
the summation by the integral as
\begin{equation}
\label{eq:e7}
  \delta G^{(0)}(z) \simeq 
  4(\int_{0}^{E_{n_{min}}}+\int_{E_{n_{max}}}^{\infty})
  (\frac{1}{z-E}+
  \frac{E}{E^{2}+1})dE.
\end{equation}
Here, one should notice that the mean level density is constant and 
equal to one in our parameterization.
The integral in Eq.(\ref{eq:e7}) is elementary and leads to
\begin{equation}
\label{eq:e8}
  \delta G^{(0)}(z) \simeq 
  4(F(z,E_{n_{min}})-F(z,0)-F(z,E_{n_{max}})),
\end{equation}
where the function $F$ is defined by 
\begin{equation}
\label{eq:e9}
 F(z,E)=\frac{1}{2} \log \frac{E^{2}+1}{(z-E)^{2}}.
\end{equation}
If $1 \ll E_{n_{min}} < z < E_{n_{max}}$ and 
$z \simeq \frac{E_{n_{min}}+E_{n_{max}}}{2}$, then one obtains
\begin{equation}
\label{eq:e10}
  \delta G^{(0)}(z) \simeq 
  4(\log z + \log \frac{E_{n_{min}}}{E_{n_{max}}}).
\end{equation}
Notice that the first term $\log z$ comes from $F(z,0)$. 
This shows that if one evaluates, for example, $E_{1000} 
(\simeq 1000)$ according 
to the truncation (I), the numerical error
\[ \delta G^{(0)}(1000) \simeq 4(\log 1000+ \log \frac{500}{1500})
   \simeq 4(6.90-1.09) \simeq 23.2, \]
is accompanied. 
We have numerically checked the validity of this estimate 
of the error.
Also, Eq.(\ref{eq:e10}) shows that the error is much 
larger as one increases the energy. Clearly, the underestimation 
of the Green's function leads to the underestimation of the eigenvalues.

The accuracy of zeros is not directly related to the magnitude of the 
error in the Green's function, but to the ratio between the 
magnitude of the error and the derivative of the Green's function 
at the zero. 
Therefore, we further examine the derivative of the Green's function. 
As a typical example, we show in Table 1 some eigenvalues around 
$E_{1000}$ obtained by the truncation (I) 
and the derivative of the 
Green's function at the corresponding zero. 
For comparison, we show the result with the truncation (II). 
It can be easily seen from Table 1 that 
whereas the zero has a fairly good accuracy 
if the derivative there is large enough compared to the error
(about 20), 
this is not the case if the derivative is small. 
In fact, some eigenvalues have numerical errors comparable to 
the mean energy difference between the nearest neighboring levels.
It is also unfortunate for the truncation (I) that the sequence 
of the absolute value of the derivatives looks like random. 
So, the accuracy of a zero just next to a very accurate one 
can be very poor. 
This of course has a serious influence on $P(S)$.

On the contrary, the numerical error by the truncation (II) 
is given by 
\begin{equation}
\label{eq:e11}
  \delta G^{(0)}(z) \simeq -4F(z,E_{n_{max}}),
\end{equation}
and quite small even for $E_{4000} \simeq 4000$
\begin{equation}
\label{eq:e12}
  \delta G^{(0)}(4000) \simeq -\frac{4z}{E_{n_{max}}} \simeq -0.16.
\end{equation}
Also, one can see that the large magnitude of the derivatives of the 
Green's function ensures the accuracy of the zeros. 
The absolute value of the typical error with a zero 
is estimated to be at most of the order of $10^{-3}$, 
namely $0.1\%$ compared to the mean level spacing. 

The physical reason why such large phase space is necessary is obvious. 
It is the singularity of the interaction 
between the unperturbed levels. 
In fact, any pair of even-even parity states couples to each other 
with the same coupling strength, because the scatterer point is located 
at the center in the billiard. The singularity of the interaction 
is a common feature to certain kinds of the pseudointegrable 
systems. Extreme care in numerical accuracy is required 
in order to analyze such systems.

The reminiscence of the Poisson-like behavior (regularity) for $S>1$
in the strong coupling limit
is somewhat surprising. Although the reason for that is one of 
the most exciting topics, it goes beyond our present scope and 
remains as a future problem. We hope that our preliminary study 
in this paper serves as a guidepost 
leading to the right direction for studying 
the rich field of the new class of quantum chaos.

In summary, we have shown that 
the level spacing distribution $P(S)$ of 
the rectangular billiard with a point-like scatterer 
in the strong coupling limit belongs to a new class. 
Contrary to the previous conclusion, it does not show 
Wigner-like behavior, but shows Poisson-like for $S>1$, 
although there remains the level repulsion in the small $S$ region. 
A wide range of the Fourier basis is demanded
in order to get the correct eigenvalues of this system.
\\[1cm]
   One of the authors (T.S.) is grateful for the support 
by the Grand-in-Aid for Encouragement of Young Scientists 
(No.04740142) by the Ministry of Education, Science and Culture.
Numerical computations have been performed on HITAC M-880 of Computer
Centre, University of Tokyo.

\newpage
{\bf Figure Captions} \\[1cm]
Figure 1: Schematic graph of the Green's function in Eq.(\ref{eq:e3}).
\\[1cm]
Figure 2: Level spacing distribution $P(S)$ in case of the strong 
coupling limit according to the truncation (I);
$n_{min}=l-500$ and $n_{max}=l+500$ 
in Eq.(\ref{eq:e5}). 
Statistics are taken within the eigenstates indicated in the figure.
The Wigner (solid line) and Poisson (broken line) are also shown 
for reference.
\\[1cm]
Figure 3: Same as Fig.2 except that according to the truncation (II);
$n_{min}=1$ and $n_{max}=100000$ 
in Eq.(\ref{eq:e5}).
\\[1cm]

\begin{table}
\caption
{
Zeros and the derivatives of the Green's function 
at the corresponding zeros. 
The second and third columns show the results according to the
truncation (I), whereas the fourth and fifth columns in case of 
the truncation (II).}
\begin{center}
\begin{tabular}{|r|r|c|r|c|} \hline
\multicolumn{1}{|c|}{n} &
\multicolumn{1}{c|}{$E_{n}$} &
\multicolumn{1}{c|}{$|(G_{appr}^{(0)})'(E_{n})|$} &
\multicolumn{1}{c|}{$E_{n}$} &
\multicolumn{1}{c|}{$|(G_{appr}^{(0)})'(E_{n})|$} \\ \hline
 995  &  994.29  &   49    &  994.52  &   251  \\ \hline 
 996  &  995.60  &   23    &  996.03  &   169  \\ \hline 
 997  &  996.55  &   89    &  996.71  &   292  \\ \hline 
 998  &  997.30  &   95    &  997.43  &   365  \\ \hline 
 999  &  999.39  &   12    & 1000.13  &   110  \\ \hline 
1000  & 1000.46  &  782    & 1000.49  &  1032  \\ \hline 
1001  & 1001.95  &   16    & 1002.44  &   178  \\ \hline 
1002  & 1003.50  &   15    & 1004.24  &   111  \\ \hline 
1003  & 1004.75  &   77    & 1005.00  &   159  \\ \hline 
1004  & 1005.31  &  618    & 1005.34  &   873  \\ \hline 
1005  & 1006.32  &   30    & 1006.70  &   171  \\ \hline 
\end{tabular}
\end{center}
\end{table}
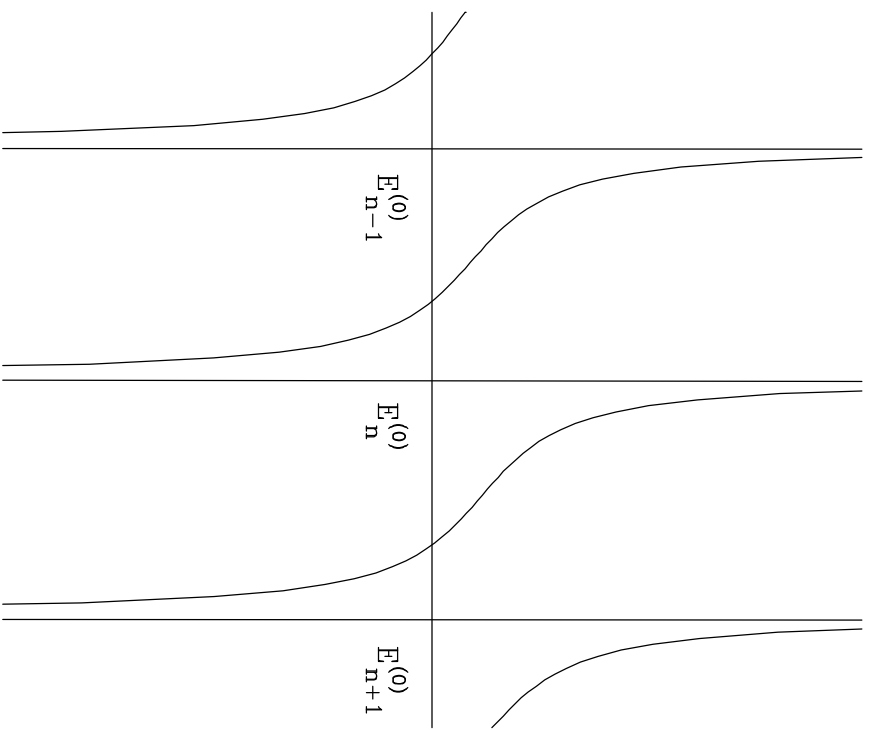
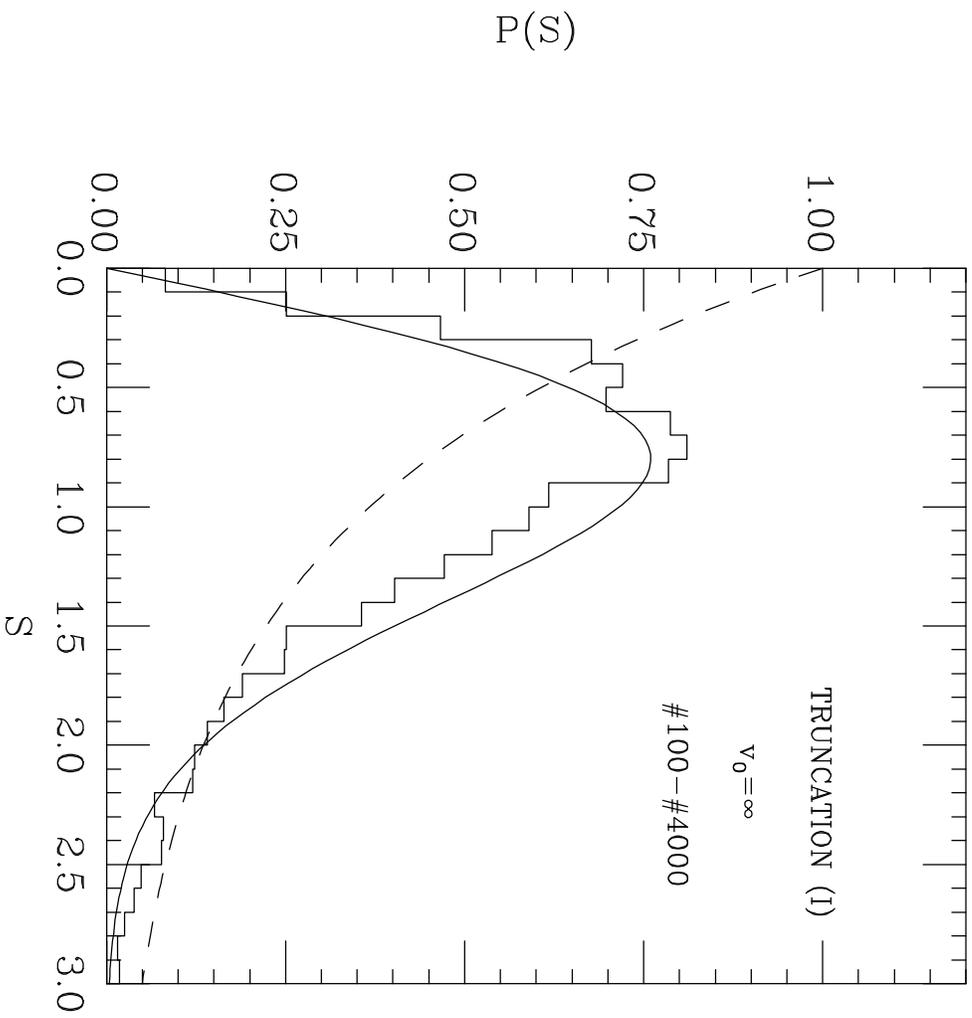
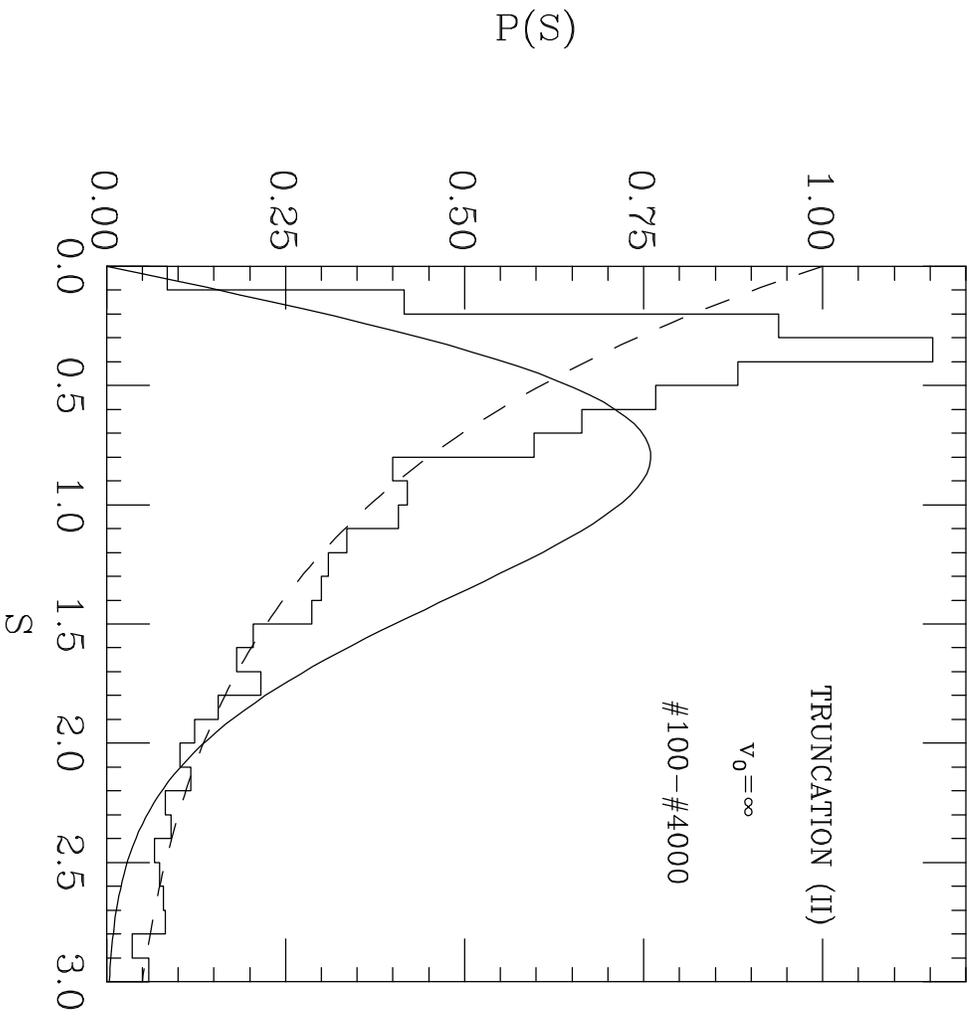
\end{document}